\documentclass[twocolumn]{aastex631}
\usepackage{amsmath}

\begin{document}

\title{High-soft to low-hard state transition in black hole X-ray binaries with GRMHD simulations}

\correspondingauthor{Indu K. Dihingia}
\email{ikd4638@sjtu.edu.cn, ikd4638@gmail.com}

\author[0000-0002-4064-0446]{Indu K. Dihingia}
\affiliation{Tsung-Dao Lee Institute, Shanghai Jiao-Tong University, Shanghai, 520 Shengrong Road, 201210, People's Republic of China}

\author[0000-0002-8131-6730]{Yosuke Mizuno}
\affiliation{Tsung-Dao Lee Institute, Shanghai Jiao-Tong University, Shanghai, 520 Shengrong Road, 201210, People's Republic of China}
\affiliation{School of Physics \& Astronomy, Shanghai Jiao-Tong University, Shanghai, 800 Dongchuan Road, 200240, People's Republic of China}
\affiliation{Institut f\"{u}r Theoretische Physik, Goethe Universit\"{a}t, Max-von-Laue-Str. 1, 60438 Frankfurt am Main, Germany}

\author[0000-0003-2635-4643]{Prateek Sharma}
\affiliation{Department of Physics, Indian Institute of Science, Bangalore, KA 560012, India}
 
\begin{abstract}
To understand the decaying phase of outbursts in the black hole (BH) X-ray binaries (BH-XRBs), we performed very long general relativistic magneto-hydrodynamic (GRMHD) simulations of a geometrically thin accretion disk around a Kerr BH with slowly rotating matter injected from outside. We thoroughly studied the flow properties, dynamical behavior of the accretion rate, magnetic flux rate, and jet properties during the temporal evolution. Due to the interaction between the thin disk and injected matter, the accretion flow near the BH goes through different phases. The sequence of phases is: soft state $\rightarrow$ soft-intermediate state $\rightarrow$ hard-intermediate state $\rightarrow$ hard state $\rightarrow$ quiescent state. 
For the accretion rate (and hence the luminosity) to decrease (as observed) in our model, the mass injection should not decay slower than angular momentum injection.
We also observed quasi-periodic oscillations (QPOs) in the accretion flow. Throughout the evolution, we observed
low-frequency QPOs ($\sim 10$Hz) and high-frequency QPOs ($\sim 200$Hz). Our simple unified accretion flow model for state transitions is able to describe outbursts in BH-XRBs.

\end{abstract}


\section{Introduction} \label{sec:intro}

The black hole (BH) X-ray binaries (BH-XRBs) are primarily dormant, with occasional outbursts triggered by a significant increase in accretion activity. In BH-XRBs, a hardness-intensity diagram, also known as the `Q' diagram, is typically used to characterize an outburst as well as the features of its various spectral states (hard state, hard/soft intermediate state, and soft state) \citep[see for detail,][]{Fender-etal2004, Remillard-McClintock2006,Done-etal2007, Dunn-etal2010,Belloni2010, Belloni-Motta2016}. 

In the soft state, the thermal component dominates the spectrum, which is explained by a geometrically thin but optically thick accretion disk in the vicinity of the BH \citep {Shakura-Sunyaev1973, Novikov-Thorne1973}. A power-law component dominates the spectrum in the hard state, and on the contrary, the contributions from both thermal and
power-law components can be seen in the intermediate state \citep[e.g.,][]{Homan-Belloni2005, McClintock-etal2006, Belloni2010}. The power-law component is caused by the Compton upscattering of soft X-ray photons due to the hot electrons present in the corona close to the BH \citep{Thorne-Price1975, Sunyaev-Truemper1979, Chakrabarti-Titarchuk1995}.
 
In the last few decades, the
consensus has been that the combination of hot accretion flow and standard thin-disk in different degrees can explain different states of the outburst quite nicely \citep[e.g.,][]{Narayan-Yi1994,Narayan-Yi1995, Esin-etal1997, Narayan-McClintock2012, Chakrabarti-Titarchuk1995,Chakrabarti2018}. However, it is still unknown how the different states are connected and how or why they transition from one to another. Realistic simulations of accretion flow may help understand the state transitions. 
Nonetheless, simulation for a realistic timescale (hours to tens of days) of an outburst is challenging, especially using general relativistic magneto-hydrodynamics (GRMHD). The framework of GRMHD is indispensable for studying physics around BHs \citep[see,][references therein]{Moscibrodzka2019,Davis-Tchekhovskoy2020, Mizuno2022,Dexter-etal2021,Dihingia-etal2022}.

In this work, we report a unified time evolution model for the decaying phase of the outburst in BH-XRBs by GRMHD simulations.
Our model can show continuous transitions of spectral states from one to another in the decaying part of the outburst. 
In the decaying phase, we can safely neglect the contribution of radiative cooling to the structure of the accretion flow. 
Radiative cooling is more important in triggering the formation of the soft state (SSD) in the rising phase \citep[see][]{Meyer-etal2007,Das-Sharma2013,Wu-etal2016,Liu-Qiao2022}. 

In section \ref{sec:math}, we briefly discuss the numerical setup, and in the following sections (\ref{sec:result0},\ref{sec:result1}), we discuss the results obtained from our simulations. Finally, in section \ref{sec:summary}, we summarise our results and discuss future plans. 

\section{Numerical setup}\label{sec:math}

This work performs a set of axisymmetric (2D) ideal GRMHD simulations using the GRMHD code \texttt{BHAC} \citep{Porth-etal2017, Olivares-etal2019} in Modified Kerr-Schild coordinates. We apply a spherical polar grid $(r, \theta, \phi)$ where the grid spacing is logarithmic in the radial direction and linear in the polar direction. The simulations use a generalized unit system with $G=M_{\rm BH}=c=1$, where $G$, $M_{\rm BH}$, and $c$ are the universal gravitational constant, the mass of the BH, and the speed of light, respectively. The distance and time are in units of $r_g=GM_{\rm BH}/c^2$ and $t_g=GM_{\rm BH}/c^3$, respectively.
Here simulations are performed considering $M_{\rm BH}=10M_{\odot}$, for that, the simulation time unit $t_g$ corresponds to $4.91\times10^{-5}$ sec.

We set up an unmagnetized geometrically thin disk in
rotational equilibrium around a Kerr BH (spin parameter $a=0.94$) initially \citep{Dihingia-etal2021,Dihingia-Vaidya2022a}. It is based on the standard thin disk model by \cite{Novikov-Thorne1973}. The density distribution of the thin disk in Boyer-Lindquist coordinates ($r,\theta$) is given by,
\begin{equation}
\begin{aligned}
\rho(r,\theta) = \rho_e(r)\times~~~~~~~~~~~~~~~~~~~~~~~~~~~~~~~~~~~~~~~~~~~~~~ &\\
    \begin{cases}
    \exp\left(-\frac{\alpha^2 z^2}{H^2}\right)\,,& \text{for $r<r_{\rm out}$}\\ 
    \exp\left(-\frac{\alpha^2 z^2}{H^2}\right) \exp\left(-(r/r_{\rm out} - 1)^2\right)\,, &
    \text{otherwise}\,,
    \end{cases}
\end{aligned}
\label{eq-rho}
\end{equation}
where $z=r\cos(\theta)$. The density distribution is kept in a thin disk configuration by choosing $\alpha=2$, and an appropriate disk height profile ($H$) following \citet{Riffert-Herold1995} and \citet{Peitz-Appl1997}. Here $\rho_e(r)$ is the density profile of the standard thin disk in the equatorial plane (Eq. 7 of \cite{Dihingia-etal2021}). As seen in Eq.~\ref{eq-rho}, the initial thin disk is exponentially decaying to an external medium at $r_{\rm out}$ ($=30r_g$, fixed). The external medium is set by following floor treatment for GRMHD (e.g., \cite{Porth-etal2017}).
For the initial velocity profiles of the thin disk, we set $u^r=0, u^\theta=0$ and provide initial azimuthal velocity $u^\phi_{\rm Kep}$ following \cite{Dihingia-etal2021} (see Eq.~12 of the reference).

We consider that the accretion flow is subjected to radiative cooling following \cite{Dihingia-etal2023,Dihingia-etal2023a}. However, for simplicity, we do not consider electron thermodynamics in this study. 
Accordingly, we apply the single temperature approximation $(T=m_pc^2 \Theta/k_{\rm B}, \Theta=p/\rho)$ to calculate the radiative cooling rates. We consider Bremsstrahlung ($Q_{\rm br}$) and synchrotron ($Q_{\rm cs}$) radiative cooling processes to be dominated in the optically thin regime.
With this, the total radiative cooling rate is calculated as:
\begin{align}
    Q_{\rm thin}^-=Q_{\rm br} + \eta Q_{\rm cs}\,,
    \label{eq:thincool}
\end{align}
where the Compton enhancement factor $(\eta)$, which is calculated following \citep{Narayan-Yi1995}. We do not provide the explicit expression of the cooling rates to avoid repetition. The detailed expression can be found in \cite{Esin-etal1996, Dihingia-etal2023,Dihingia-etal2023a}.

In a thin disk, the optically thick components of the radiative losses can not be neglected due to the high optical depth. Therefore, we consider 
generalized cooling formula capable of choosing radiative cooling processes depending on the optical depth as suggested by \cite{Narayan-Yi1995} and \cite{Esin-etal1996},
\begin{align}
    Q^{-}_{\rm tot} = \frac{4\sigma_T T^4/H}{1.5\tau + \sqrt{3} + \tau_{\rm abs}^{-1}}\,,
    \label{eq:totq}
\end{align}
where $\tau = \tau_{\rm es} + \tau_{\rm abs}$, $\tau_{\rm es}=2\sigma_Tn_e H$, and $\tau_{\rm abs}=(HQ_{\rm thin}^-/\sigma_T T^4)$. Here, $H$ is the scale height of the accretion flow, which is obtained as $H=T^4/|\nabla T^4|$ \citep{Fragile-Meier2009}. Finally, $n_e$ is the number density of the electrons, and $\sigma_T$ is the Thomson cross-section of the electron.

Along with the thin disk around the BH, we inject matter continuously from the region away from the BH with a radius $r=r_{\rm inject}$. The properties of the injected material are considered as
\begin{equation}
\begin{aligned}
    &\rho_{\rm in}(r,\theta,t) = \rho_{\rm in}(r,\theta)\exp\left(-t/t_{\rm scale, \rho}\right); \\
    &u^r=0,~ u^\theta=0, ~{\rm and}~u^\phi=f_0 u^\phi_{\rm Kep}\exp\left(-t/t_{\rm scale, u^\phi}\right)\,,
\end{aligned}
\label{eq:02}
\end{equation}
where, $t_{\rm scale, \rho}$ and $t_{\rm scale, u^\phi}$ define the timescales of the injection for density and angular velocity, and $f_0$ ($=0.4$, fixed for current work to facilitate sub-Keplerian injection of matter) denotes the fraction of the ratio of the azimuthal velocity of injected matter with respect to the Keplerian velocity at the given radius. 
For this pilot study, we use $t_{\rm scale, u^\phi} = 50,000\,t_g = 2.455\,M_{\rm BH}/10\,M_\odot$ sec, and choose three values of $t_{\rm scale, \rho}= 50,000\,t_g, 75,000\,t_g$, and $100,000\,t_g$, i.e., $2.455, 3.68,$ and 4.91\,$M_{\rm BH}/10\,M_\odot$ sec. We label them as MOD1, MOD2, and MOD3, respectively. We perform a simulation of MOD1 with a higher resolution (MOD1HR). Furthermore,
we choose, $\rho_{\rm in}$ as, 
\begin{equation}
\rho_{\rm in}=0.1\rho_e(r)\exp\left(-\frac{\alpha_{\rm in}^2 z^2}{H^2}\right)\exp\left(-\left(\frac{r-r_{\rm inject}}{8}\right)^2\right).
\label{eq-rhoin}
\end{equation}
For the current study, we choose $\alpha_{\rm in}=8$ and $r_{\rm inject}=400\,r_{g}$. We consider the injected material to be magnetized, and for simplicity, it contains a poloidal single-loop magnetic field which is given by the following vector potential,
\begin{equation}
    A_\phi \propto \max(\rho_{\rm in}/\rho_{\rm max}-10^{-5},0).
    \label{eq-aphi}
\end{equation}
The strength of the initial magnetic field is set by a minimum value of plasma-$\beta$ ($=p_{\rm gas}/p_{\rm mag}$) at the injection region as $\beta_{\rm in} = p_{\rm gas}^{\rm in}/p_{\rm mag}^{\rm in}|_{\rm min}=200$. Note that the choice of the parameters for this study is not unique. We expect similar qualitative results for different sets of parameters in similar ranges. We discuss the time evolution of the initial setup without the injected matter and only with injected matter (without the thin disk) in Appendix A, which justifies our choice of initial setup and the parameters.

The simulation domain is $r \in [0.96\,r_g, 1000\,r_g]$ and $\theta \in [0,\pi]$. We perform two simulations with different maximum effective resolutions of $640\times288$ (base resolution $320\times 144$; MOD1-3) and $1280\times640$ (base resolution $640\times 288$; MOD1HR). To resolve the equatorial thin disk better, the maximum resolution is concentrated near the equator within $\theta=\pi/2.4$ to $\theta=\pi - \pi/2.4$. Simulations run up to time $t=10^6\,t_g=49.1\,M_{\rm BH}/10M_\odot$ sec. Since we need to evolve our simulations for a long time, even our higher resolution simulation cannot afford to adequately resolve the magnetorotational instability (MRI) in the later time due to the weakening of the magnetic field strength over time. However, in the initial phase, we resolved the fastest-growing MRI wavelength, at least with quality factor $Q_\theta\gtrsim6$ for $t\lesssim t_{\rm scale,u^\phi}$ at the disk surface, where magnetized injected matter and unmagnetized thin-disk interacts.
\section{Flow structure}\label{sec:result0}
\begin{figure*}
    \centering
    \includegraphics[scale=0.55]{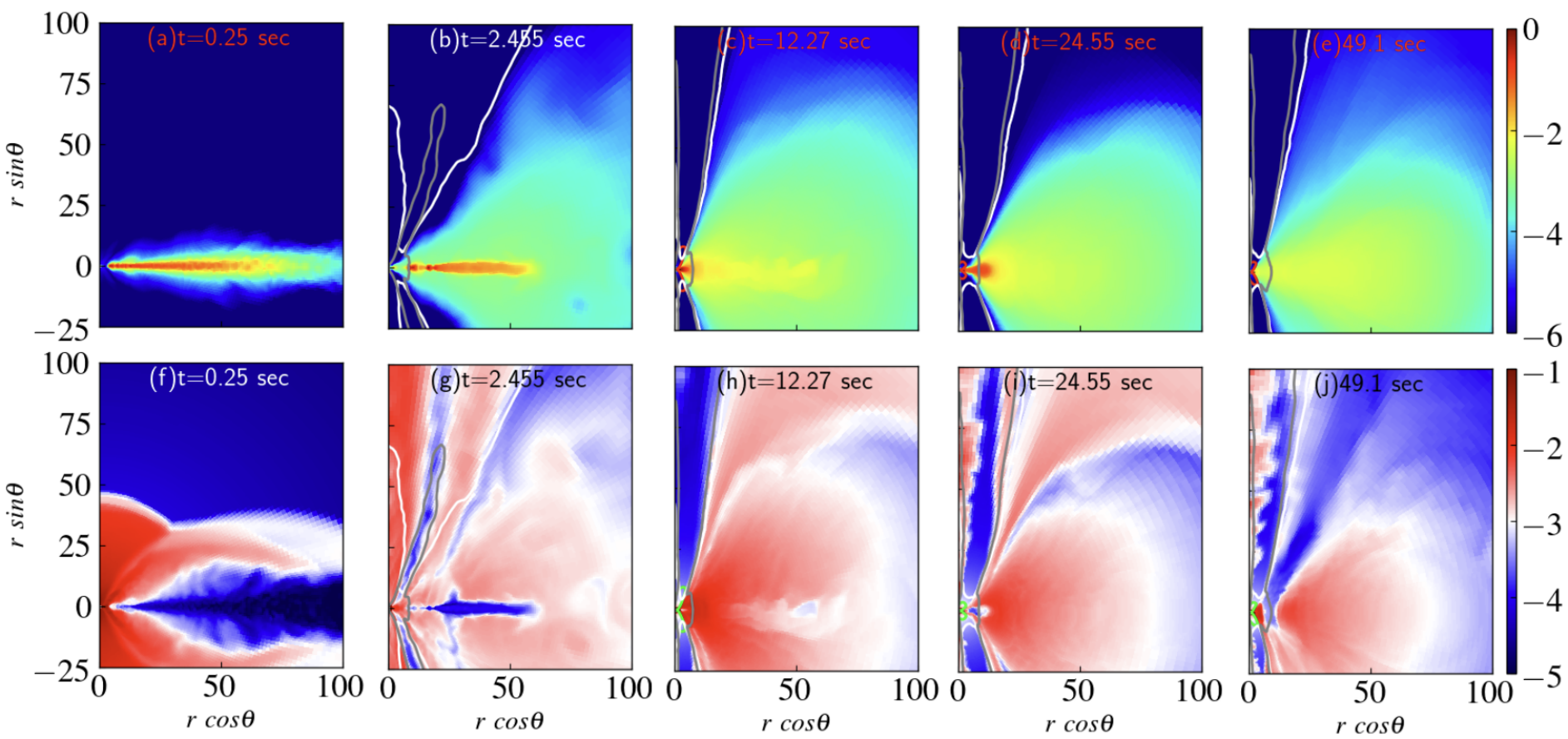}
    \caption{Logarithmic time-average density (upper panels) and temperature (lower panels) distribution at different times marked on the top of the figures. The time averaging is performed within $0.1$ sec around the marked time. The white, red (upper)/green (lower), and grey solid lines show $-hu_t=1$, $\sigma=1$, and $\gamma=1.1$ contours.}
    \label{fig:denisty}
\end{figure*}
In this section, we discuss the accretion flow structures in different phases of evolution for MOD1. At time, $t=0.25, 2.455, 12.27, 24.55,$ and $49.1$ $M_{\rm BH}/10\,M_\odot$ sec, we plot the time-averaged density and temperature $(p/\rho)$ distribution in Fig.~\ref{fig:denisty}. The time average is performed within $0.1$ sec around the marked time. The white red (upper)/green (lower), and grey solid lines in the figure show the Bernoulli parameter $-hu_t=1$, magnetization $\sigma=b^2/\rho=1$, and Lorentz factor $\gamma=1.1$ contours. In addition to these panels, follow the supplementary material, where we show a continuous transition of flow structure up to the end of the simulation for MOD1.

Initially, we started with a cold (low temperature) and geometrically thin accretion disk around the BH, which can be seen in panels (a and f) of Fig.~\ref{fig:denisty}. In this phase, the matter of the disk has Keplerian angular momentum. 
From the start of simulations, the injected matter interacts with the thin disk, creating a hot corona surrounding the thin disk. The matter in the hot corona has lower angular momentum than the high-density thin disk. We show the density and temperature snapshots at this time range in panels (b and g) of Fig.~\ref{fig:denisty}. 
This suggests a Keplerian thin disk sandwiched by sub-Keplerian components on both sides of the equatorial plane.  
Such a structure of accretion flow is usually identified as a two-component accretion flow (TCAF). In such a flow, most of the radiation comes from the thin disk at the equatorial plane. However, due to the hot corona surrounding the thin disk, some hard X-rays are expected to be observed by the inverse-Compton process. Therefore, such a TCAF may explain the soft-intermediate states of an outburst \citep[][and references therein]{Chakrabarti-Titarchuk1995, Chakrabarti2018}. Unlike the pure-thin disk case, in this stage, we see outflows from the disk (see $-hu_t>1$ and $\gamma>1.1$  region around the polar region). In this phase, a weak jet is launched. This weak jet region becomes stronger during the transition to the next phase of the evolution (see section~\ref{sec:result1} for detail). 

With further evolution, the thin disk loses its angular momentum due to the onset of magnetized wind, and matter in the accretion flow heats up. The wind redistributes the high-angular momentum matter far from the BH and creates a very large torus (see panels (c and h) of Fig.~\ref{fig:denisty}). It suggests that the density of the flow in most of the region is very low compared to the earlier two phases (panels (a) and (b) of Fig.~\ref{fig:denisty}). On the contrary, the temperature of the flow in most of the region is very high compared to the earlier two phases (panels (f) and (g) of Fig.~\ref{fig:denisty}).
These facts strongly suggest a flow behavior  
in the hot accretion flow regime (e.g., advection-dominated accretion flow (ADAF)). In this phase, we see a well-evolved steady jet, which is more compact and collimated than in the earlier stages (see the $-hu_t>1$ and $\sigma=1$ contours in the panel). 
In this phase, the whole accretion flow acts as a hot corona around the BH. Therefore, we expect the dominant radiation component to be mostly hard X-rays produced from synchrotron self-Comptonization and Comptonization of bremsstrahlung photons. However, the optical depth of the flow close to the BH may be slightly higher due to the large size of the torus. This may contribute to some soft X-ray components in the spectra. Accordingly, this phase may be associated with the hard-intermediate state of an outburst.

After creating a large torus, the matter in the torus keeps reducing due to accretion, and the torus becomes smaller with time.
We show the corresponding density and temperature distribution in panels (d and i) of Fig.~\ref{fig:denisty}. The maximum density in the torus decreases by two orders of magnitudes, which keeps reducing at the end of simulation time.

The reduction of torus size and density decreases the effective optical depth of the flow close to the BH. Accordingly, we may expect an abundance of hard X-rays in the emission spectra. However, the total luminosity will start to decrease as the total mass of the accretion flow drops with time. In this phase, the jet becomes much narrower than in earlier stages (see the $-hu_t>1$ and $\gamma=1.1$ contours in the panel). This phase resembles a radiatively inefficient quiescent state (RIAF) \citep[e.g.,][]{Narayan-etal1996, Narayan-etal1997, Hameury-etal1997}. Thus, we observe that the accretion flow makes a smooth transition from a cold (SSD) disk to a hot torus structure during the evolution.  

\section{Time series analysis}\label{sec:result1}
\begin{figure*}
    \centering
    \includegraphics[scale=0.48]{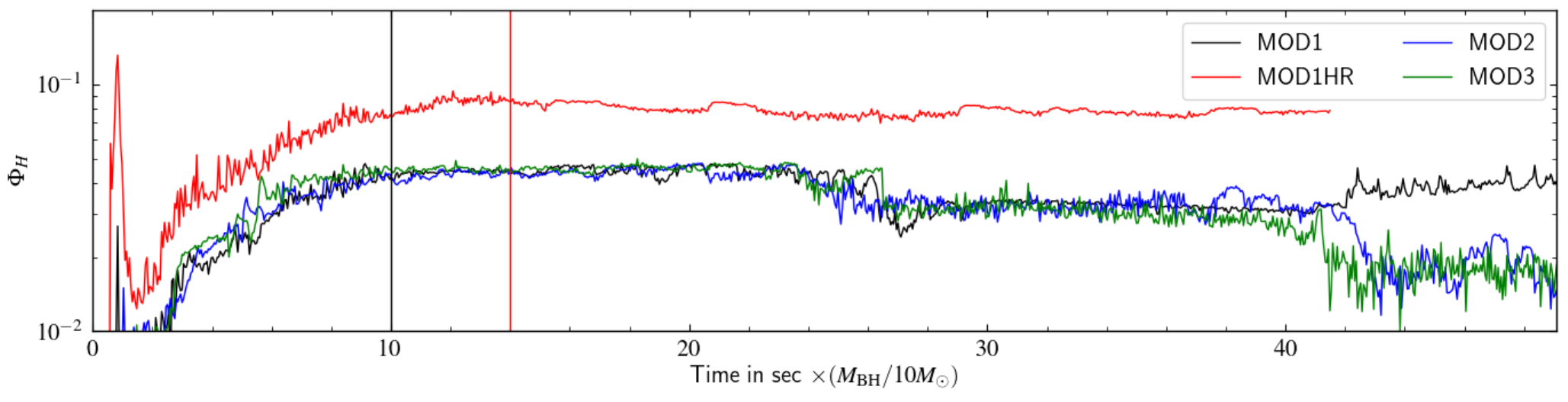}
    \caption{Evolution of the magnetic flux at the event horizon $(\Phi_H)$ as a function of time. The black, red, blue, and green lines correspond to models MOD1, MOD1HR, MOD2, and MOD3, respectively  
    The corresponding vertical lines indicate the transition from MDHAF to IDLAF.}
    \label{fig:mag_flx}
\end{figure*}

\begin{figure*}
    \centering
    \includegraphics[scale=0.65]{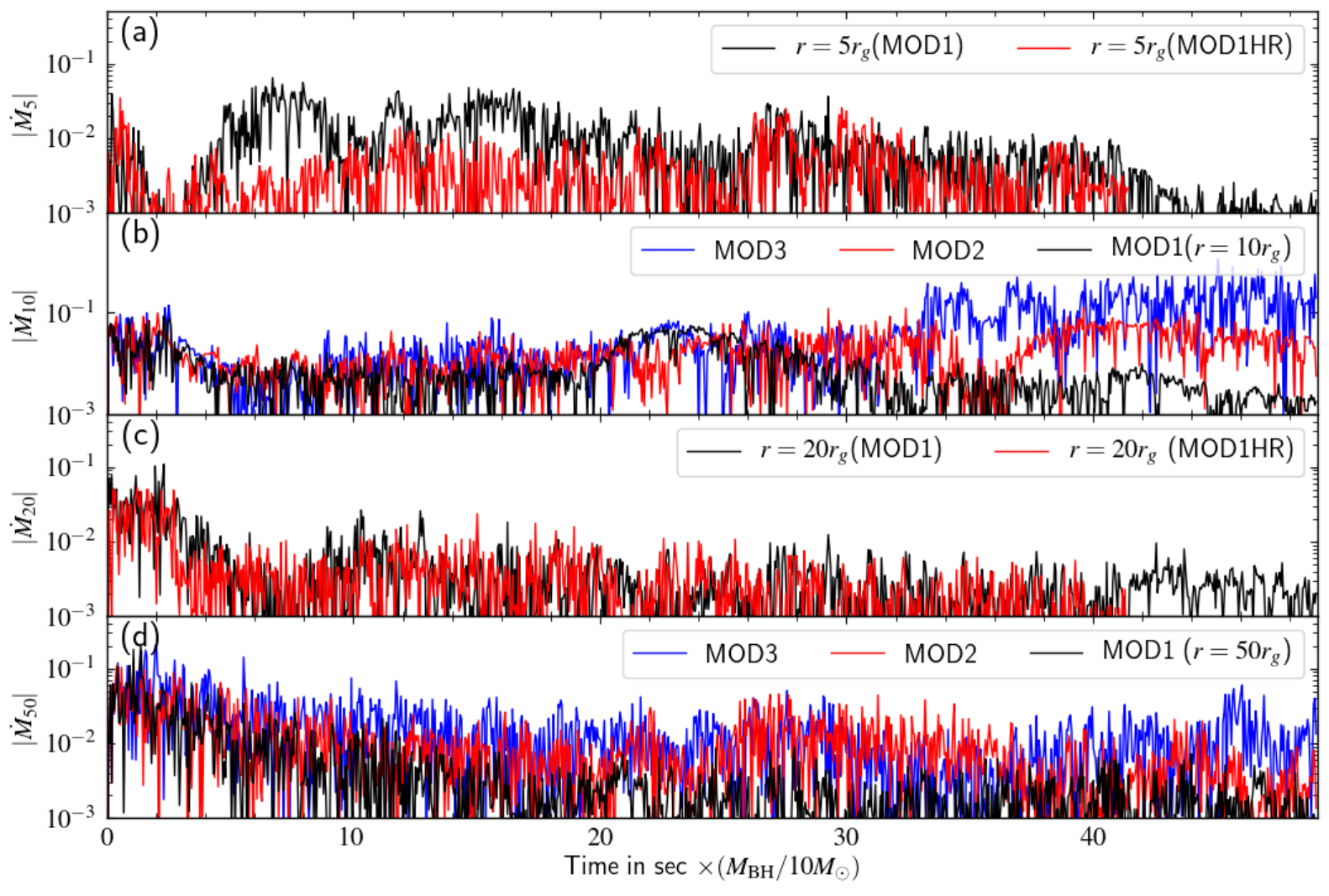}
    \caption{
    Evolution of the accretion rate at different radii $r=5r_g,10r_g,20r_g$, and $50r_g$ from top to bottom.
    In panels (a) and (c), a comparison is shown between MOD1 and MOD1HR, and in panels (b) and (d), a comparison between MOD1, MOD2, and MOD3 is shown.}
    \label{fig:rates_ini}
\end{figure*}

Along with the flow structure, it is also important to study the time series
from the runs to characterize different states further. Following this,  Fig.~\ref{fig:mag_flx} presents the time evolution of magnetic flux ($\Phi_{H}$) at the horizon for MOD1 (black), MOD1HR (red), MOD2 (blue), and MOD3 (green), (for detail definition, see Eq.~(96) of \cite{Porth-etal2017}).

We can classify the evolution of the flow into magnetic-driven high-angular momentum flow (MDHAF) and infall-driven low-angular momentum flow (IDLAF) phases by looking at these evolutions. 
In the MDHAF, high angular momentum is transported outwards due to active magnetic winds (e.g., \cite{Dihingia-etal2021}), causing accretion towards the BH. During this phase, the magnetic flux increases with time. For MOD1, we observe that up to $t \sim 10\,M_{\rm BH}/10\,M_\odot$ sec (see the black vertical line), the magnetic flux increases. The maximum magnetic flux observed for MOD1 is $\Phi_{\rm H}\sim 0.04$. After that, magnetic flux remains roughly constant, which we identify as IDLAF. However, we observe slight drops in the magnetic flux within time $t \sim 25-40\,M_{\rm BH}/10\,M_\odot$ sec.

By comparing the magnetic flux 
evolution for two different numerical resolution cases, MOD1 and MOD1HR, we find that they follow a similar increasing trend initially and remain roughly constant after time $t > 14\,M_{\rm BH}/10\,M_\odot$ sec. The transition is shown by the vertical red line in the figure. However, we see distinct quantitative differences between the values of the magnetic flux at both resolutions. The maximum magnetic flux for MOD1HR is about $\Phi_{\rm H}\sim 0.08$. Accordingly, we expect a stronger relativistic jet in MOD1HR as compared to MOD1. We discuss this in the next section in detail. Interestingly, unlike the low-resolution model, we do not observe any drops in between the evolution.

Comparing 
simulation models with different $t_{\rm scale, \rho}=50,000, 75,000, 100,000\,t_g$ (MOD1, MOD2, MOD3) in Fig.~\ref{fig:mag_flx}, we find that magnetic flux profiles evolve slightly differently for them. Although they show the transition from MDHAF to IDLAF in exactly a similar manner and at the same location, i.e., $t\sim10\,M_{\rm BH}/10\,M_\odot$ sec. For a higher value of $t_{\rm scale, \rho}$, more low-angular momentum matter is supplied from the injection radius. It facilitates faster mixing of injected matter and the high angular momentum matter in the thin disk. Interestingly, we note that the magnetic flux drops significantly with time after $t>25\,M_{\rm BH}/10\,M_\odot$ sec for MOD2 and MOD3. The magnetic flux at the end of the simulation is lower for a higher value of $t_{\rm scale, \rho}$. The average value of magnetic flux for MOD1, MOD2, and MOD3, at the end of simulation $(t=10^6\,t_g)$ is about $\Phi_{\rm H}\sim 0.040, 0.019$, and $0.017$, respectively. This suggests that $t_{\rm scale, \rho}$ plays a role in deciding the characteristics throughout the simulation as well as the final stage.

Next, we would like to study the accretion rate as a function of time. To do that, we plot the absolute value accretion rate for MOD1 at different radii $|\dot{M}_5| (r=5r_g), |\dot{M}_{10}| (r=10r_g), |\dot{M}_{20}| (r=20r_g)$, and $|\dot{M}_{50}| (r=50r_g)$, in panels Fig.~\ref{fig:rates_ini}(a)-(d), respectively. In panels Fig.~\ref{fig:rates_ini}(a) and (c), we also plot the same
evolution for the high-resolution run (MOD1HR) for comparison. In Fig.~\ref{fig:rates_ini}(b) and (d), we also plot the accretion rate
evolution for different values of $t_\rho$, i.e., for models MOD2 and MOD3 for comparison. 
For MOD1, we observe that for higher radii $(r=20,50r_g)$, the accretion rate always shows a decreasing trend. However, at radius $r=10$, the accretion rate at $t=20-25$$\,M_{\rm BH}/10\,M_\odot$ sec shows an increasing trend, but after that, it drops monotonically. On the contrary, at a smaller radius $(r=5r_g)$, the initial accretion rate is smaller but the trend is mostly the same.

Ideally, the accretion rate in the thin disk is supposed to be very high
due to the viscosity present in the thin disk.
Such viscosity could be triggered in a thin disk due to magneto-rotational instabilities (MRI) if a weak magnetic field is present in the flow \citep{Balbus-Hawley1998}. We performed some 3D test simulations with a magnetic field and presented the results in Appendix B. In these simulations, we observed three orders of magnitude higher accretion rate in the thin disk with resolved MRI as compared to unresolved MRI. The numerical cost of such simulations is extremely high. On top of that, we require simulations of temporal length $t\sim10^6t_g$, which is impossible to achieve with current numerical resources. Apart from this caveat, axisymmetric consideration is the most efficient framework to perform longer simulations. Accordingly,  we neglect any such viscosity of the flow and continue with axisymmetric consideration in this paper.
Accordingly, we observe that the accretion rate 
close to the black hole is smaller at the beginning of the simulation because it is in
equilibrium. 

Note that, during the transition from MDHAF to IDLAF, the accretion flow makes a transition from a cold thin disk to a hot thick disk (torus) (see Fig. \ref{fig:denisty}). Accordingly, the accretion rate of the thin disk due to the viscosity should also vanish (not studied here). Although Fig.~\ref{fig:rates_ini}(a-b) shows a slightly different shape of the accretion rate 
evolution with time, 
it is expected to be half of the bell-shaped curve
after the accretion of the thin disk component, with its maximum at the beginning. Such accretion rate 
evolution can be clearly seen at radii $r=20,50r_g$ (see Fig.~\ref{fig:rates_ini}(c-d)). This behavior of the accretion rate
evolution is expected for the accretion rate in the decaying phase of an outburst\citep{Esin-etal1997, Ferreira-etal2006, Kylafis-Belloni2015}.

By comparing the
evolution of model MOD1 (black) and MOD1HR (red) in Fig.~\ref{fig:rates_ini}(a-c), we observe that they behave mostly similarly with some quantitative differences. The primary driver of accretion in our simulation is the interaction between unmagnetized high-angular momentum and magnetized low-angular momentum matter. This process creates the magnetized wind, which transports angular momentum and leads to accretion in the low-resolution models. Apart from this process, the high-resolution model (MOD1HR) subjected to resolved MRI at least $t\lesssim t_{\rm scale,u^\phi}$. 
Due to that, the high-resolution model accumulates higher magnetic flux, which consequently results in stronger winds from the thin disk compared to the low-resolution model. Notably, these differences are most prominent near the black hole (see Fig.~\ref{fig:rates_ini}a). However, farther from the black hole, the differences in the 
evolution become negligible (see Fig.~\ref{fig:rates_ini}c). Additionally, as the matter in the thin disk depletes over time, the disparity between the accretion  
evolution of MOD1 and MOD1HR diminishes, even close to the black hole.  

Finally, we want to compare the accretion rate 
evolution for different simulation models with different $t_{\rm scale, \rho}$. As we mentioned earlier, a higher value of $t_{\rm scale, \rho}$ ensures more injected low angular momentum. In panel Fig.~\ref{fig:rates_ini}(d), we observe that the accretion rate for all the models shows an overall decrease with time. However, for the higher value of $t_{\rm scale, \rho}$ (MOD3), the decrease in accretion rate is slower as compared to the lower value of $t_{\rm scale, \rho}$ (MOD1). On the contrary, in Fig.~\ref{fig:rates_ini}(b), the accretion rate at $r=10r_g$ for MOD2 shows a more or less constant profile and slightly decreases towards the end, unlike MOD1, where the accretion rate shows a 
roughly decreasing tendency throughout. However, in model MOD3 with the highest value of $t_{\rm scale, \rho}$, the accretion 
rate evolution shows an increasing trend with time. It is needless to mention that with more time the accretion rate will certainly decrease with time. We
observe a decaying trend of accretion rate in the decaying phase of an outburst \citep{Esin-etal1997, Ferreira-etal2006, Kylafis-Belloni2015}. Therefore, it implies that higher values of $t_{\rm scale, \rho}$ than those of $t_{\rm scale, u_\phi}$, which leads to an increase in accretion rate (and hence luminosity), are not preferable to explain commonly known features of outburst in BH-XRBs. The equal decaying time-scale for angular momentum ($t_{\rm scale, u_\phi}$) and mass ($t_{\rm scale, \rho}$) is preferred. Accordingly, in the next two sections, we will only focus on MOD1 and MOD1HR for the analysis.
\begin{figure*}
    \centering
    \includegraphics[scale=0.45]{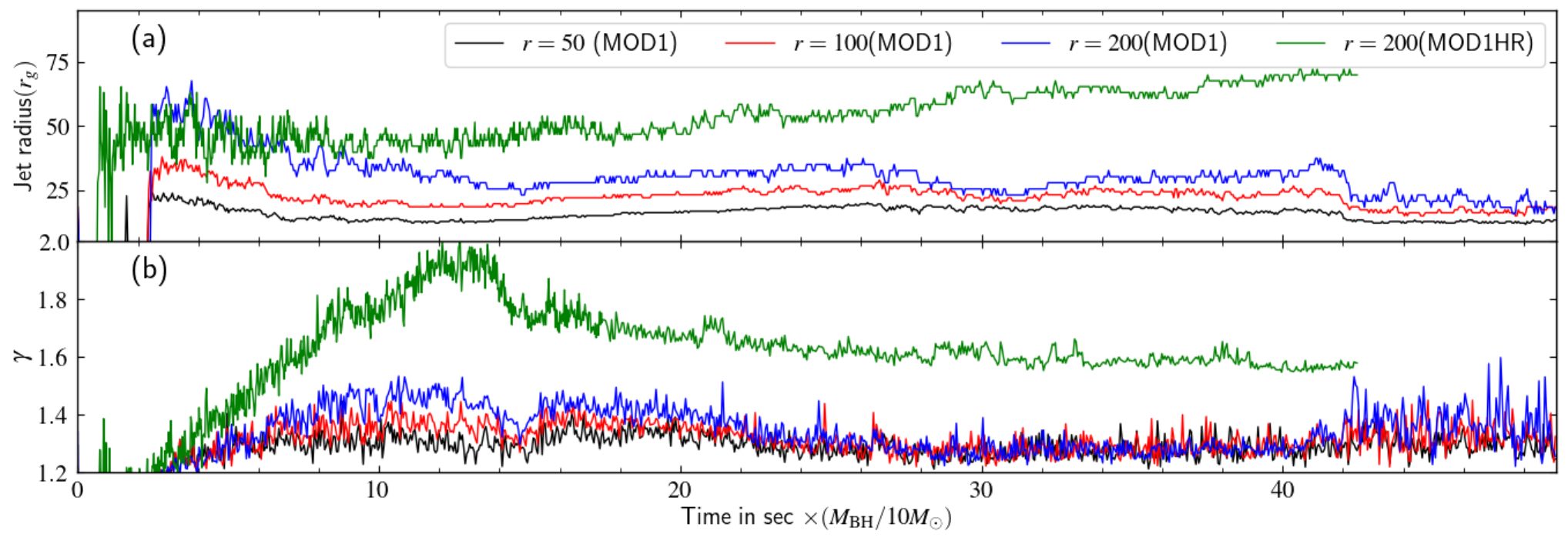} 
    \caption{Evolution of jet radius and the Lorentz factor $(\gamma)$ as a function of time. The black, red, and blue lines in the figure refer to the quantities calculated at radius $r=50,100$, and $200r_g$. The green line corresponds to the high-resolution run.}
    \label{fig:jet-prop}
\end{figure*}
%
\subsection{Jet properties}

Next, we discuss the properties of the jet in terms of the jet radius and the Lorentz factor. We consider any outflow with a Lorentz factor $\gamma>1.1$ (or jet velocity $v_j\sim 0.42\,c$) as a jet for simplicity. With this definition, Fig.~\ref{fig:jet-prop} shows jet radius and Lorentz factor ($\gamma$) as a function of time for MOD1 and MOD1HR. The black, red, and blue lines in the figure refer to the quantities calculated at radius $r=50$, $100$, and $200\,r_g$, respectively, for MOD1 model. Subsequently, we also plot the same quantities for the high-resolution (MOD1HR) run at $r=200\,r_g$ in green for comparison. We first explain, the jet properties for MOD1, and after that we will compare that with MOD1HR.

We do not see any jets initially because the magnetic flux accumulated around the BH is not high enough to launch a jet. After $t>2\,M_{\rm BH}/10\,M_\odot$ sec, we start to see a weak but very broad jet when magnetic flux starts to accumulate on the BH (see panel (c) of Fig.~\ref{fig:rates_ini}). 
With time, the jet radius monotonically decreases, suggesting the jet becomes more narrow and compact with time. Although, within time $t\sim 10-42 \,M_{\rm BH}/10\,M_\odot$ sec, we observe a slight increase in the jet radius, but eventually the jet radius drops significantly after $t>42 \,M_{\rm BH}/10\,M_\odot$ sec. 
However, the Lorentz factor does not show a monotonic behavior with time. It roughly follows
the magnetic flux 
evolution as shown in Fig.~\ref{fig:mag_flx}. Initially, the Lorentz factor increases with the increase of magnetic flux, giving a maximum of $\gamma\sim 1.5$ for MOD1 at $r=200r_g$. After flow changes its nature from MDHAF to IDLAF ($t\sim 10 8\,M_{\rm BH}/10\,M_\odot$ sec), the Lorentz factor starts to decrease with a minimum value $\gamma\sim 1.3$ for MOD1 at all radii. After $t > 42\,M_{\rm BH}/10\,M_\odot$ sec, the Lorentz factor starts to increase again but in an episodic fashion. The jet radius keeps decreasing with time. Note that the compact jet is expected in the radiatively inefficient hard state \citep[see for discussion][]{Kylafis-etal2012}.

Comparing the results from high-resolution (MOD1HR) and low-resolution runs (MOD1) in Fig.~\ref{fig:jet-prop}, we see pretty similar behavior in the Lorentz factor. In the higher resolution run, the magnetorotational instabilities (MRI) could be resolved marginally, which helps in stronger disk winds than that of MOD1. This affects the maximum magnetic flux regions, where we observe the strongest jet. Accordingly, the maximum Lorentz factor for MOD1HR is about $\gamma\sim1.9$, which is much larger than that of the maximum value for model MOD1 ($\gamma\sim1.4$). This shows that the jet physics depends on the resolution of the simulation, and therefore, higher resolution runs are necessary to capture physics related to the jet. On the contrary, the jet radius initially starts with exactly similar evolution, but after some time, we observe a higher increasing trend in comparison to the low-resolution run (blue line). This is due to the fact that the numerical viscosity reduces with resolution. Accordingly, the size of the torus reduced slowly in comparison with the low-resolution model (MOD1). Also, magnetic flux is much higher for MOD1HR, which can expand the jet with time due to the reduction of the supports from the sheath region. Eventually, we expect the jet radius to decrease with time after a sufficiently long simulation time. 

\subsection{Oscillations in the accretion flow}
\begin{figure*}
    \centering
    \includegraphics[scale=0.48]{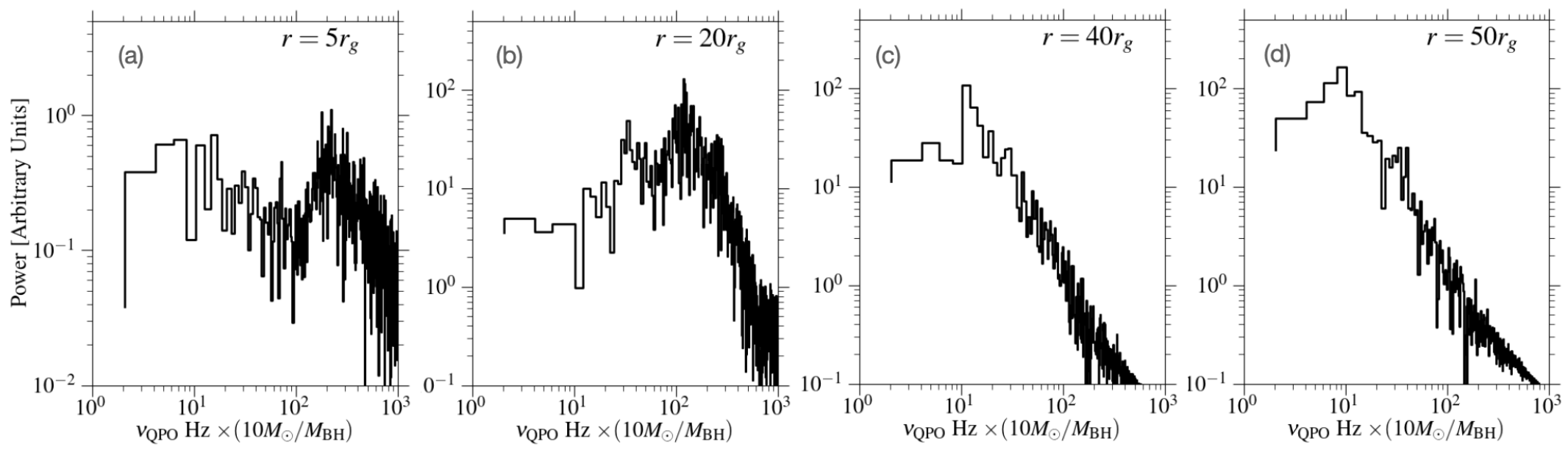}
    \caption{Plots of PDS of accretion rates at different radii marked on each panel within simulation time $t=2-2.5\, M_{\rm BH}/10 M_\odot$ sec.}
    \label{fig:PDS1}
\end{figure*}

Due to the instabilities between infalling matter and the outflow, the accretion flow exhibits some oscillations.
To analyze the oscillations in more detail, we calculate the power density spectra (PDS) of accretion rate at different radii $r=5r_g,20r_g,40r_g$ and $50r_g$ for MOD1 at simulation time $t=2-2.5 M_{\rm BH}/10 M_\odot$ sec and plot them in Fig.~\ref{fig:PDS1}, where the corresponding times are written on each panel in units of $M_{\rm BH}/10 M_\odot$. At this time, the PDS shows clear peaks in all the radii, suggesting the existence of quasi-periodic oscillations (QPO) in the accretion flow. However, the frequency of oscillations $(\nu_{\rm QPO})$ differs at different radii. The value of the $\nu_{\rm QPO}$ is about $\sim 200, 100, 10$, and $8$Hz $\times 10M_\odot/M_{\rm BH}$ at radii $r=5,20,40,$ and $50r_g$, respectively. Note that the frequencies of oscillation are neither equal to the local Keplerian frequency nor equal to radial epicyclic frequency. Although, at radii $r=40$ and $50r_g$, they are quite close to the Keplerian frequency ($\sim $ radial epicyclic frequency for $r\gg1$). 
However, the frequency of oscillation changes with time at a given radius.

\begin{figure*}
    \centering
    \includegraphics[scale=0.5]{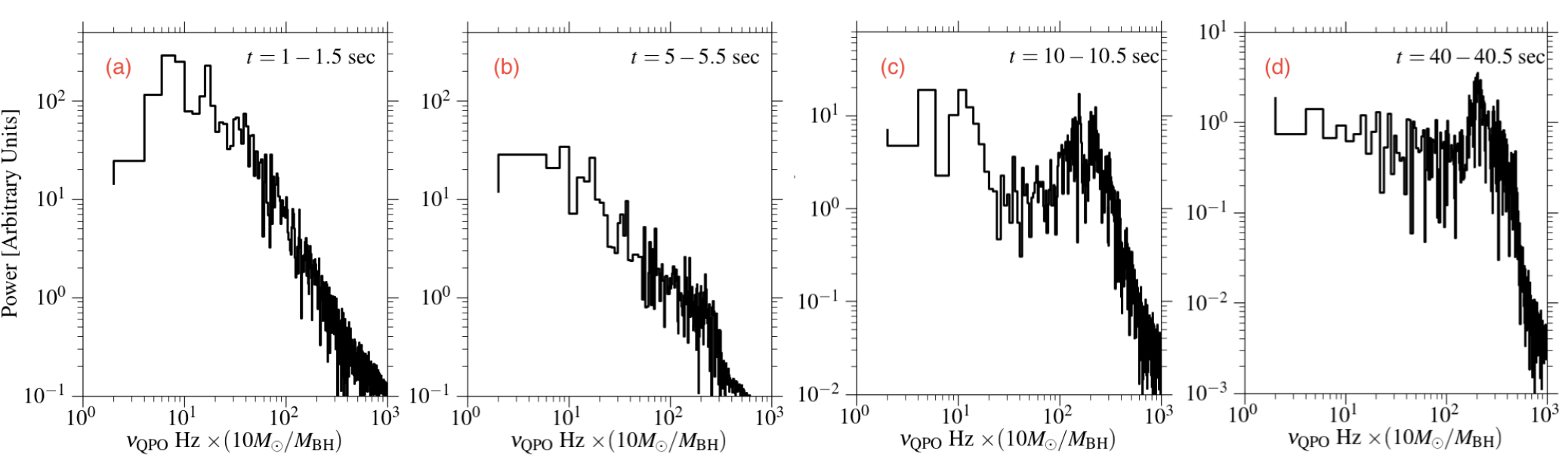}
    \caption{Plots of PDS of $\dot{M}_{50}$ 
    calculated over different times, which are marked on the panels.}
    \label{fig:PDS}
\end{figure*}

Next, we calculate the PDS at different simulation times for MOD1 and we plot them in Fig.~\ref{fig:PDS}.
For this figure, we only calculate the PDS of accretion rate radius $r=50$ ($\dot{M}_{50}$). We observe that the shape of the PDS changes over time as the flow changes its characteristics. The frequency of oscillation also changes with time.
These QPO oscillations in the accretion flow can modulate the emission from the accretion flow, resulting in the observed QPOs in BH-XRBs. Note that, in this phase of evolution, the observed QPO frequency will depend on the observed energy of the emission and the dominant region (radii) of that emission range of the disk.

\begin{figure*}
    \centering
    \includegraphics[scale=0.55]{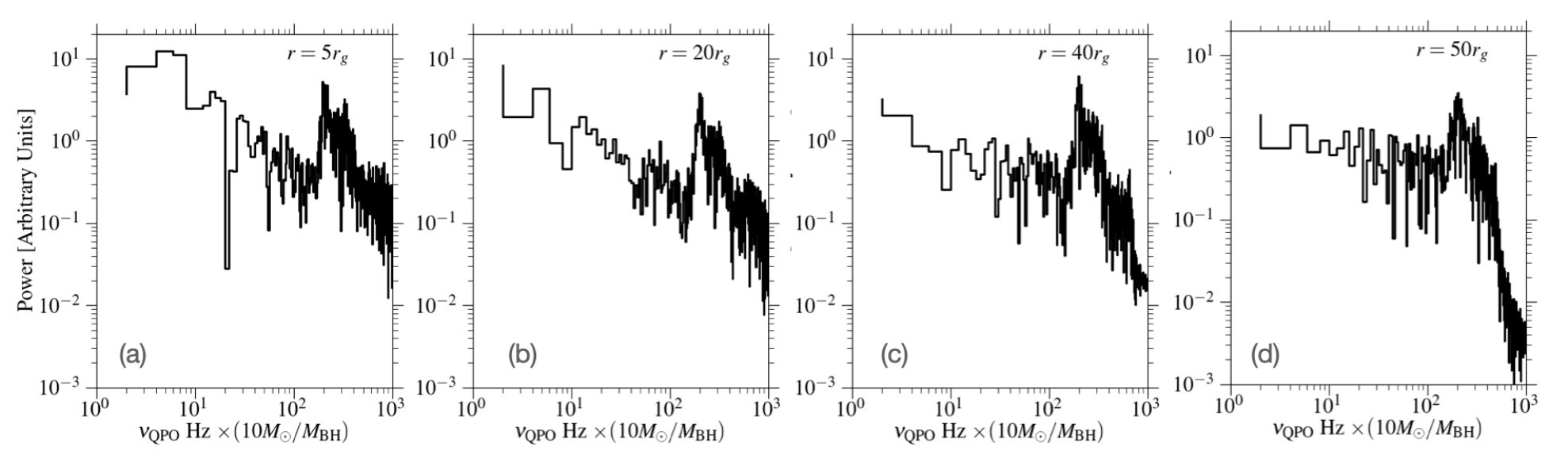}
    \caption{Plots of PDS of accretion rates at different radii marked on each panel within simulation time $t=40-40.5\,M_{\rm BH}/10 M_\odot$ sec.}
    \label{fig:PDS2}
\end{figure*}
By comparing panels in Fig.~\ref{fig:PDS} and Fig.~\ref{fig:PDS1}(d), we find that in early simulation time up to $t\sim10\,M_{\rm BH}/10\,M_\odot$ sec, we observe sharp peaks with QPO frequencies in the range of $\nu_{QPO} \sim 10 \times 10\,M_\odot/M_{\rm BH}\,\mathrm{Hz}$. 
For example, in the panels (Figs.~\ref{fig:PDS}a, \ref{fig:PDS1}d, \ref{fig:PDS}b-c) at time $t=1-1.5$, $2-2.5$, $5-5.5$, and $10-10.5\,M_{\rm BH}/10\,M_\odot$ sec, the obtained QPO frequencies are $\nu_{QPO} \sim 6$, $8$, $8$, and $\sim 10 \times 10\,M_\odot/M_{\rm BH}\,\mathrm{Hz}$, respectively. 
At time $t=10-10.5\,M_{\rm BH}/10\,M_\odot$ sec, we also observe a QPO frequency at $\nu_{QPO}\sim200 \times 10\,M_\odot/M_{\rm BH}\,\mathrm{Hz}$. However, in panel Fig.~\ref{fig:PDS}d at time $t=40-40.5\,M_{\rm BH}/10\,M_\odot$ sec, we only observe a prominent peak at high frequency, $\nu_{QPO}\sim200\times10\,M_\odot/M_{\rm BH}\,\mathrm{Hz}$. 

The variation of QPO frequencies with time indicates that the frequency of oscillation depends on the injected angular momentum of the flow. During this phase, the ratio of injected angular momentum to the Keplerian angular momentum at the innermost stable circular orbit (ISCO) $\lambda_{\rm inj}/\lambda_{\rm Kep}(ISCO)\gg1$, we observe low-frequency QPOs ($\sim 10\,\mathrm{Hz}$). However, during the phase $\lambda_{\rm inj}/\lambda_{\rm Kep}(ISCO)\sim1$, we do not observe prominent QPO features. 
Finally, for the phase with $\lambda_{\rm inj}/\lambda_{\rm Kep}(ISCO)\ll1$, we observe high-frequency QPOs ($\sim 200\,\mathrm{Hz}$)  in the accretion flow. Accordingly, after a sufficiently long simulation time, we expect a similar frequency of oscillation at different radii. To show that, we plot PDS calculated in a similar way as Fig.~\ref{fig:PDS1} but at later time $t=40-40.5\,M_{\rm BH}/10\,M_\odot$ sec in Fig.~\ref{fig:PDS2}. All the panels of the figure show prominent $\nu_{QPO}\sim200 \times 10\,M_\odot/M_{\rm BH}\,\mathrm{Hz}$. This essentially also suggests that the structure of the accretion flow in a low-angular momentum would be a torus structure at this time. 
The density maximum of the torus at simulation time $t\sim40\,M_{\rm BH}/10\,M_\odot$ sec is around $r_{\rm max}\sim6r_g$, and the Keplerian frequency of that radius is around $\nu_{\rm Kep}\sim200 \times 10\,M_\odot/M_{\rm BH}\,\mathrm{Hz}$, which show good resemblance with the $\nu_{QPO}$ of the PDS at all radii in Fig.~\ref{fig:PDS2}.

\begin{figure*}
    \centering
    \includegraphics[scale=0.55]{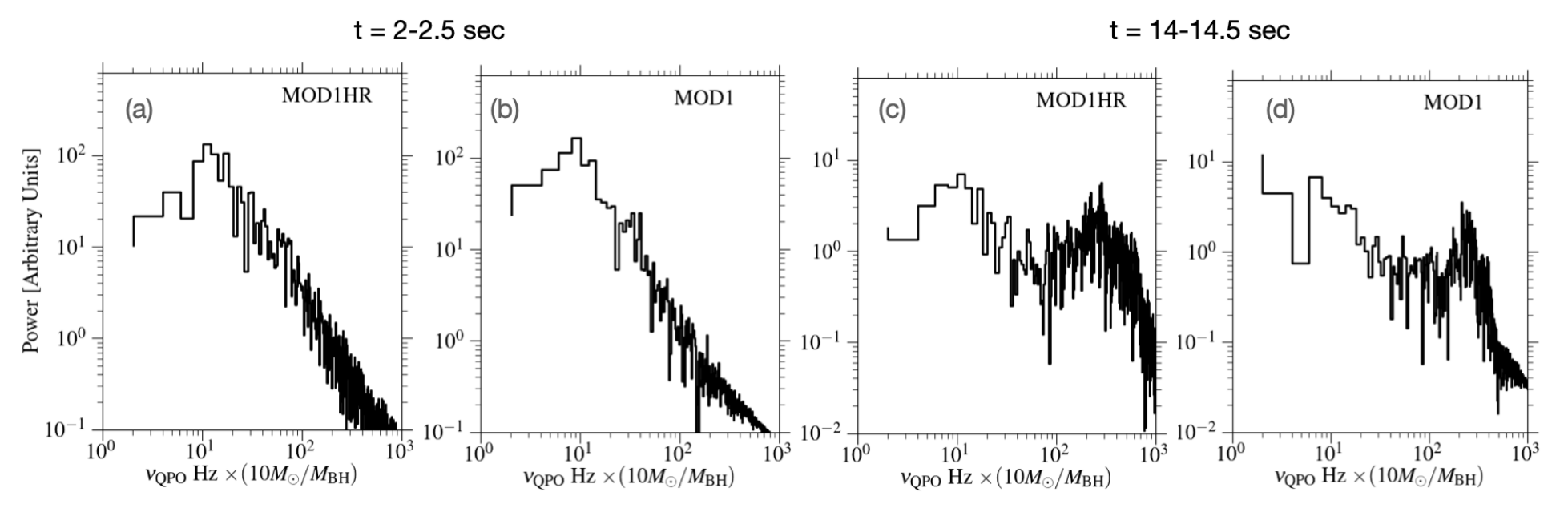}
    \caption{Comparison of PDS calculated for accretion rates $\dot{M}_{50}$ for MOD1 and MOD1HR marked for two different time $t=2-2.5$ (a,b) and $14-14.5\,$(c,d)$M_{\rm BH}/10 M_\odot$ sec.}
    \label{fig:PDS3}
\end{figure*}

Finally, we would like to see the impact of high resolution for consistency. To do that, we plot PDS at time $t=2-2.5\,M_{\rm BH}/10\,M_\odot$ sec Fig.~\ref{fig:PDS3}(a,b) and at time $t=14-14.5 \,M_{\rm BH}/10\,M_\odot$ sec (c,d) for model MOD1HR and MOD1 for comparison. The PDS shows excellent consistency in both resolutions. This result also suggests that the production of QPOs in our simulation does not depend on unresolved/partially resolved MRI. On the contrary, the observed QPO would be due to the mixing of low-angular momentum and high-angular momentum matter, which we expect to act similarly in both resolutions.
\section{Summary and discussion}\label{sec:summary}

In summary, as the injected matter interacts with the thin disk around the BH, the thin cold disk transitions to a two-component accretion flow. With time, more matter is injected from the outer edge. As a result, the accretion flow becomes a geometrically thick hot torus. Eventually, the torus size decreases, and finally, it is left with very low-density hot matter around the BH. Although we have not calculated the spectra throughout the evolution, still looking at the density, temperature, magnetization, accretion rate 
evolution, and jet properties, we can comment on the possible spectral evolution. 

Throughout our simulation, flow transitions as follows: soft state $\rightarrow$ soft-intermediate state $\rightarrow$ hard-intermediate state $\rightarrow$ hard state $\rightarrow$ quiescent state, which exactly resembles the decaying stage of an outburst. Quite fascinatingly, we could detect the jet properties as expected by many observations, and they suggest jets only in the intermediate states \citep[see][]{Belloni-etal2011, Kylafis-etal2012, Kylafis-etal2018}. Also, we observed a distinct variation in QPO frequency throughout the evolution and observed both kinds of frequencies, i.e., low-frequency QPOs ($\sim 10$Hz) and high-frequency QPOs ($\sim 200$Hz). 

Based on our knowledge, there are no such reports in the literature showing continuous transitions between spectral states with a unified model. However, it is always anticipated that the accretion rate will take a bell-shaped nature with time for an outburst \citep[e.g.,][]{Esin-etal1997,Ferreira-etal2006,Kylafis-Belloni2015, Kumar-Yuan2022}. For our simulation, we only supply half of the bell-shaped curve for the injected matter at the outer edge, following our interest in simulating only the decaying phase. Earlier, \cite{Das-Sharma2013}, have also shown with hydrodynamic simulations that the transition from a thin disk to a RIAF happens because of mass exhaustion due to accretion (with different injection models and without the jets and QPO signatures). This suggests that the decaying injected matter is the only essential ingredient to trigger the decaying phase of an outburst, which is very generic in nature, and its qualitative features are independent of the detailed modeling.

A longer decaying time scale for mass as compared to angular momentum results in an increasing accretion rate close to the black hole, which is not expected for commonly known decaying phase outburst. Accordingly, we suggest that an equal decaying time scale for mass and angular momentum is more appropriate to explain the decaying phase of commonly known outbursts in BH-XRBs.

Additionally, our high-resolution run shows quite similar results for accretion rates and QPO frequencies. However, for the high-resolution case, we observe a higher accumulation of magnetic flux around the event horizon. Thus, we observe stronger jets in the high-resolution run than that in the low-resolution run. This essentially suggests that to handle the physics of the jet precisely, the setup that MRI is well-resolved will be better. On top of that, the asymmetric nature of our simulation may also impact the jet physics significantly. It is well known that asymmetric MRI can not sustain dynamo for the long term (e.g.,anti-dynamo theorem, \cite{Cowling1933}). Accordingly, a full 3D simulation would be much better for understanding the jet. However, performing such a very long, highly resolved 3D run is extremely challenging with current computational resources. A mean-field dynamo approach will be the most efficient way to include in asymmetric calculations, where it is possible to accomplish 3D dynamo effects \citep[e.g.,][]{Parker1955, Krause1980, Sadowski-etal2015}. Nonetheless, with 3D or with a mean-field dynamo, we expect that the morphology of the Lorentz factor or the magnetic flux evolution remains unaltered. However, we expect some changes in the maximum value of the Lorentz factor and the jet radius obtained from the current results.

Finally, we want to mention some additional caveats and the future studies we plan as a follow-up to this work. In this work, we only demonstrate the decaying stage of an outburst with GRMHD simulations. It will be in our priorities to simulate the full cycle of an outburst (including the rising phase). Note that a realistic outburst in BH-XRBs could be of a few hours to months long and rarely even years  \cite[for e.g.,][]{Yan-Yu2015}, which is impossible to perform under the current framework. Therefore, more quantitative and parameter space surveys are required to correlate the realistic decay time with the physical properties of injected matter ($\beta_{\rm in}, f_0, t_{\rm scale, \rho}$), the initial thin accretion disk ($r_{\rm out}$, initial thickness), or the properties of the central BH ($a, M_{\rm BH}$, etc.). Moreover, throughout our study, we consider a single temperature approximation, which overestimates the temperature of electrons. Therefore, a robust two-temperature approach would be worth trying for self-consistent results \citep[e.g.,][]{Sadowski-etal2017, Dihingia-etal2023}. However, we expect that the key features of our study will not alter with it.

\begin{acknowledgments}
This research is supported by the National Natural Science Foundation of China (Grant No. 12273022) and the Shanghai Municipality orientation program of basic research for international scientists (grant no. 22JC1410600). PS acknowledges a Swarnajayanti Fellowship from the Department of Science and Technology, India (DST/SJF/PSA- 03/2016-17).
The simulations were performed on Pi2.0 and Siyuan Mark-I at Shanghai Jiao Tong University.
This work has made use of NASA's Astrophysics Data System (ADS). We appreciate the thoroughness and thoughtful comments provided by the reviewers that have improved the manuscript.
\end{acknowledgments}

\bibliography{references}{}

\begin{thebibliography}{}
\expandafter\ifx\csname natexlab\endcsname\relax\def\natexlab#1{#1}\fi
\providecommand{\url}[1]{\href{#1}{#1}}
\providecommand{\dodoi}[1]{doi:~\href{http://doi.org/#1}{\nolinkurl{#1}}}
\providecommand{\doeprint}[1]{\href{http://ascl.net/#1}{\nolinkurl{http://ascl.net/#1}}}
\providecommand{\doarXiv}[1]{\href{https://arxiv.org/abs/#1}{\nolinkurl{https://arxiv.org/abs/#1}}}

\bibitem[{{Balbus} \& {Hawley}(1998)}]{Balbus-Hawley1998}
{Balbus}, S.~A., \& {Hawley}, J.~F. 1998, Reviews of Modern Physics, 70, 1,
  \dodoi{10.1103/RevModPhys.70.1}

\bibitem[{{Belloni}(2010)}]{Belloni2010}
{Belloni}, T.~M. 2010, in Lecture Notes in Physics, Berlin Springer Verlag, ed.
  T.~{Belloni}, Vol. 794, 53, \dodoi{10.1007/978-3-540-76937-8\_3}

\bibitem[{{Belloni} \& {Motta}(2016)}]{Belloni-Motta2016}
{Belloni}, T.~M., \& {Motta}, S.~E. 2016, {Transient Black Hole Binaries}, ed.
  C.~{Bambi}, Vol. 440, 61, \dodoi{10.1007/978-3-662-52859-4_2}

\bibitem[{{Belloni} {et~al.}(2011){Belloni}, {Motta}, \&
  {Mu{\~n}oz-Darias}}]{Belloni-etal2011}
{Belloni}, T.~M., {Motta}, S.~E., \& {Mu{\~n}oz-Darias}, T. 2011, Bulletin of
  the Astronomical Society of India, 39, 409.
\newblock \doarXiv{1109.3388}

\bibitem[{{Chakrabarti} \& {Titarchuk}(1995)}]{Chakrabarti-Titarchuk1995}
{Chakrabarti}, S., \& {Titarchuk}, L.~G. 1995, \apj, 455, 623,
  \dodoi{10.1086/176610}

\bibitem[{{Chakrabarti}(2018)}]{Chakrabarti2018}
{Chakrabarti}, S.~K. 2018, in Fourteenth Marcel Grossmann Meeting - MG14, ed.
  M.~{Bianchi}, R.~T. {Jansen}, \& R.~{Ruffini}, 369--384,
  \dodoi{10.1142/9789813226609_0020}

\bibitem[{{Cowling}(1933)}]{Cowling1933}
{Cowling}, T.~G. 1933, \mnras, 94, 39, \dodoi{10.1093/mnras/94.1.39}

\bibitem[{{Das} \& {Sharma}(2013)}]{Das-Sharma2013}
{Das}, U., \& {Sharma}, P. 2013, \mnras, 435, 2431,
  \dodoi{10.1093/mnras/stt1452}

\bibitem[{{Davis} \& {Tchekhovskoy}(2020)}]{Davis-Tchekhovskoy2020}
{Davis}, S.~W., \& {Tchekhovskoy}, A. 2020, \araa, 58, annurev,
  \dodoi{10.1146/annurev-astro-081817-051905}

\bibitem[{{Dexter} {et~al.}(2021){Dexter}, {Scepi}, \&
  {Begelman}}]{Dexter-etal2021}
{Dexter}, J., {Scepi}, N., \& {Begelman}, M.~C. 2021, \apjl, 919, L20,
  \dodoi{10.3847/2041-8213/ac2608}

\bibitem[{{Dihingia} {et~al.}(2023{\natexlab{a}}){Dihingia}, {Mizuno}, {Fromm},
  \& {Rezzolla}}]{Dihingia-etal2023}
{Dihingia}, I.~K., {Mizuno}, Y., {Fromm}, C.~M., \& {Rezzolla}, L.
  2023{\natexlab{a}}, \mnras, 518, 405, \dodoi{10.1093/mnras/stac3165}

\bibitem[{{Dihingia} {et~al.}(2023{\natexlab{b}}){Dihingia}, {Mizuno}, {Fromm},
  \& {Younsi}}]{Dihingia-etal2023a}
{Dihingia}, I.~K., {Mizuno}, Y., {Fromm}, C.~M., \& {Younsi}, Z.
  2023{\natexlab{b}}, arXiv e-prints, arXiv:2305.09698,
  \dodoi{10.48550/arXiv.2305.09698}

\bibitem[{{Dihingia} \& {Vaidya}(2022)}]{Dihingia-Vaidya2022a}
{Dihingia}, I.~K., \& {Vaidya}, B. 2022, Journal of Astrophysics and Astronomy,
  43, 23, \dodoi{10.1007/s12036-022-09804-z}

\bibitem[{{Dihingia} {et~al.}(2021){Dihingia}, {Vaidya}, \&
  {Fendt}}]{Dihingia-etal2021}
{Dihingia}, I.~K., {Vaidya}, B., \& {Fendt}, C. 2021, \mnras, 505, 3596,
  \dodoi{10.1093/mnras/stab1512}

\bibitem[{{Dihingia} {et~al.}(2022){Dihingia}, {Vaidya}, \&
  {Fendt}}]{Dihingia-etal2022}
---. 2022, \mnras, 517, 5032, \dodoi{10.1093/mnras/stac3021}

\bibitem[{{Done} {et~al.}(2007){Done}, {Gierli{\'n}ski}, \&
  {Kubota}}]{Done-etal2007}
{Done}, C., {Gierli{\'n}ski}, M., \& {Kubota}, A. 2007, \aapr, 15, 1,
  \dodoi{10.1007/s00159-007-0006-1}

\bibitem[{{Dunn} {et~al.}(2010){Dunn}, {Fender}, {K{\"o}rding}, {Belloni}, \&
  {Cabanac}}]{Dunn-etal2010}
{Dunn}, R.~J.~H., {Fender}, R.~P., {K{\"o}rding}, E.~G., {Belloni}, T., \&
  {Cabanac}, C. 2010, \mnras, 403, 61, \dodoi{10.1111/j.1365-2966.2010.16114.x}

\bibitem[{{Esin} {et~al.}(1997){Esin}, {McClintock}, \&
  {Narayan}}]{Esin-etal1997}
{Esin}, A.~A., {McClintock}, J.~E., \& {Narayan}, R. 1997, \apj, 489, 865,
  \dodoi{10.1086/304829}

\bibitem[{{Esin} {et~al.}(1996){Esin}, {Narayan}, {Ostriker}, \&
  {Yi}}]{Esin-etal1996}
{Esin}, A.~A., {Narayan}, R., {Ostriker}, E., \& {Yi}, I. 1996, \apj, 465, 312,
  \dodoi{10.1086/177421}

\bibitem[{{Fender} {et~al.}(2004){Fender}, {Belloni}, \&
  {Gallo}}]{Fender-etal2004}
{Fender}, R.~P., {Belloni}, T.~M., \& {Gallo}, E. 2004, \mnras, 355, 1105,
  \dodoi{10.1111/j.1365-2966.2004.08384.x}

\bibitem[{{Ferreira} {et~al.}(2006){Ferreira}, {Petrucci}, {Henri},
  {Saug{\'e}}, \& {Pelletier}}]{Ferreira-etal2006}
{Ferreira}, J., {Petrucci}, P.~O., {Henri}, G., {Saug{\'e}}, L., \&
  {Pelletier}, G. 2006, \aap, 447, 813, \dodoi{10.1051/0004-6361:20052689}

\bibitem[{{Fragile} \& {Meier}(2009)}]{Fragile-Meier2009}
{Fragile}, P.~C., \& {Meier}, D.~L. 2009, \apj, 693, 771,
  \dodoi{10.1088/0004-637X/693/1/771}

\bibitem[{{Hameury} {et~al.}(1997){Hameury}, {Lasota}, {McClintock}, \&
  {Narayan}}]{Hameury-etal1997}
{Hameury}, J.~M., {Lasota}, J.~P., {McClintock}, J.~E., \& {Narayan}, R. 1997,
  \apj, 489, 234, \dodoi{10.1086/304780}

\bibitem[{{Homan} \& {Belloni}(2005)}]{Homan-Belloni2005}
{Homan}, J., \& {Belloni}, T. 2005, \apss, 300, 107,
  \dodoi{10.1007/s10509-005-1197-4}

\bibitem[{{Krause} \& {Raedler}(1980)}]{Krause1980}
{Krause}, F., \& {Raedler}, K.~H. 1980, {Mean-field magnetohydrodynamics and
  dynamo theory}

\bibitem[{{Kumar} \& {Yuan}(2022)}]{Kumar-Yuan2022}
{Kumar}, R., \& {Yuan}, Y.-F. 2022, arXiv e-prints, arXiv:2210.00683.
\newblock \doarXiv{2210.00683}

\bibitem[{{Kylafis} \& {Belloni}(2015)}]{Kylafis-Belloni2015}
{Kylafis}, N.~D., \& {Belloni}, T.~M. 2015, \aap, 574, A133,
  \dodoi{10.1051/0004-6361/201425106}

\bibitem[{{Kylafis} {et~al.}(2012){Kylafis}, {Contopoulos}, {Kazanas}, \&
  {Christodoulou}}]{Kylafis-etal2012}
{Kylafis}, N.~D., {Contopoulos}, I., {Kazanas}, D., \& {Christodoulou}, D.~M.
  2012, \aap, 538, A5, \dodoi{10.1051/0004-6361/201117052}

\bibitem[{{Kylafis} \& {Reig}(2018)}]{Kylafis-etal2018}
{Kylafis}, N.~D., \& {Reig}, P. 2018, \aap, 614, L5,
  \dodoi{10.1051/0004-6361/201833339}

\bibitem[{{Liu} \& {Qiao}(2022)}]{Liu-Qiao2022}
{Liu}, B.~F., \& {Qiao}, E. 2022, iScience, 25, 103544,
  \dodoi{10.1016/j.isci.2021.103544}

\bibitem[{{McClintock} {et~al.}(2006){McClintock}, {Shafee}, {Narayan},
  {Remillard}, {Davis}, \& {Li}}]{McClintock-etal2006}
{McClintock}, J.~E., {Shafee}, R., {Narayan}, R., {et~al.} 2006, \apj, 652,
  518, \dodoi{10.1086/508457}

\bibitem[{{Meyer} {et~al.}(2007){Meyer}, {Liu}, \&
  {Meyer-Hofmeister}}]{Meyer-etal2007}
{Meyer}, F., {Liu}, B.~F., \& {Meyer-Hofmeister}, E. 2007, \aap, 463, 1,
  \dodoi{10.1051/0004-6361:20066203}

\bibitem[{{Mizuno}(2022)}]{Mizuno2022}
{Mizuno}, Y. 2022, Universe, 8, 85, \dodoi{10.3390/universe8020085}

\bibitem[{{Mo{\'s}cibrodzka}(2019)}]{Moscibrodzka2019}
{Mo{\'s}cibrodzka}, M. 2019, \mnras, 490, 5353, \dodoi{10.1093/mnras/stz2875}

\bibitem[{{Narayan} {et~al.}(1997){Narayan}, {Barret}, \&
  {McClintock}}]{Narayan-etal1997}
{Narayan}, R., {Barret}, D., \& {McClintock}, J.~E. 1997, \apj, 482, 448,
  \dodoi{10.1086/304134}

\bibitem[{{Narayan} \& {McClintock}(2012)}]{Narayan-McClintock2012}
{Narayan}, R., \& {McClintock}, J.~E. 2012, \mnras, 419, L69,
  \dodoi{10.1111/j.1745-3933.2011.01181.x}

\bibitem[{{Narayan} {et~al.}(1996){Narayan}, {McClintock}, \&
  {Yi}}]{Narayan-etal1996}
{Narayan}, R., {McClintock}, J.~E., \& {Yi}, I. 1996, \apj, 457, 821,
  \dodoi{10.1086/176777}

\bibitem[{{Narayan} \& {Yi}(1994)}]{Narayan-Yi1994}
{Narayan}, R., \& {Yi}, I. 1994, \apjl, 428, L13, \dodoi{10.1086/187381}

\bibitem[{{Narayan} \& {Yi}(1995)}]{Narayan-Yi1995}
---. 1995, \apj, 452, 710, \dodoi{10.1086/176343}

\bibitem[{{Novikov} \& {Thorne}(1973)}]{Novikov-Thorne1973}
{Novikov}, I.~D., \& {Thorne}, K.~S. 1973, in Black Holes (Les Astres Occlus),
  343--450

\bibitem[{{Olivares} {et~al.}(2019){Olivares}, {Porth}, {Davelaar}, {Most},
  {Fromm}, {Mizuno}, {Younsi}, \& {Rezzolla}}]{Olivares-etal2019}
{Olivares}, H., {Porth}, O., {Davelaar}, J., {et~al.} 2019, \aap, 629, A61,
  \dodoi{10.1051/0004-6361/201935559}

\bibitem[{{Parker}(1955)}]{Parker1955}
{Parker}, E.~N. 1955, \apj, 122, 293, \dodoi{10.1086/146087}

\bibitem[{{Peitz} \& {Appl}(1997)}]{Peitz-Appl1997}
{Peitz}, J., \& {Appl}, S. 1997, Mon. Not. Roy. Astron. Soc., 286, 681,
  \dodoi{10.1093/mnras/286.3.681}

\bibitem[{{Porth} {et~al.}(2017){Porth}, {Olivares}, {Mizuno}, {Younsi},
  {Rezzolla}, {Moscibrodzka}, {Falcke}, \& {Kramer}}]{Porth-etal2017}
{Porth}, O., {Olivares}, H., {Mizuno}, Y., {et~al.} 2017, Computational
  Astrophysics and Cosmology, 4, 1, \dodoi{10.1186/s40668-017-0020-2}

\bibitem[{{Remillard} \& {McClintock}(2006)}]{Remillard-McClintock2006}
{Remillard}, R.~A., \& {McClintock}, J.~E. 2006, \araa, 44, 49,
  \dodoi{10.1146/annurev.astro.44.051905.092532}

\bibitem[{{Riffert} \& {Herold}(1995)}]{Riffert-Herold1995}
{Riffert}, H., \& {Herold}, H. 1995, Astrophys. J., 450, 508,
  \dodoi{10.1086/176161}

\bibitem[{{Sadowski} {et~al.}(2015){Sadowski}, {Narayan}, {Tchekhovskoy},
  {Abarca}, {Zhu}, \& {McKinney}}]{Sadowski-etal2015}
{Sadowski}, A., {Narayan}, R., {Tchekhovskoy}, A., {et~al.} 2015, \mnras, 447,
  49, \dodoi{10.1093/mnras/stu2387}

\bibitem[{{S{\'{a}}dowski} {et~al.}(2017){S{\'{a}}dowski}, {Wielgus},
  {Narayan}, {Abarca}, {McKinney}, \& {Chael}}]{Sadowski-etal2017}
{S{\'{a}}dowski}, A., {Wielgus}, M., {Narayan}, R., {et~al.} 2017, \mnras, 466,
  705, \dodoi{10.1093/mnras/stw3116}

\bibitem[{{Sano} {et~al.}(2004){Sano}, {Inutsuka}, {Turner}, \&
  {Stone}}]{Sano-etal2004}
{Sano}, T., {Inutsuka}, S.-i., {Turner}, N.~J., \& {Stone}, J.~M. 2004, \apj,
  605, 321, \dodoi{10.1086/382184}

\bibitem[{{Shakura} \& {Sunyaev}(1973)}]{Shakura-Sunyaev1973}
{Shakura}, N.~I., \& {Sunyaev}, R.~A. 1973, \aap, 500, 33

\bibitem[{{Sunyaev} \& {Truemper}(1979)}]{Sunyaev-Truemper1979}
{Sunyaev}, R.~A., \& {Truemper}, J. 1979, \nat, 279, 506,
  \dodoi{10.1038/279506a0}

\bibitem[{{Thorne} \& {Price}(1975)}]{Thorne-Price1975}
{Thorne}, K.~S., \& {Price}, R.~H. 1975, \apjl, 195, L101,
  \dodoi{10.1086/181720}

\bibitem[{{Wu} {et~al.}(2016){Wu}, {Xie}, {Yuan}, \& {Gan}}]{Wu-etal2016}
{Wu}, M.-C., {Xie}, F.-G., {Yuan}, Y.-F., \& {Gan}, Z. 2016, \mnras, 459, 1543,
  \dodoi{10.1093/mnras/stw742}

\bibitem[{{Yan} \& {Yu}(2015)}]{Yan-Yu2015}
{Yan}, Z., \& {Yu}, W. 2015, \apj, 805, 87, \dodoi{10.1088/0004-637X/805/2/87}

\end{thebibliography}
\bibliographystyle{aasjournal}

\appendix
\section{Testing the initial setup}
\begin{figure}
    \centering
    \includegraphics[scale=0.5]{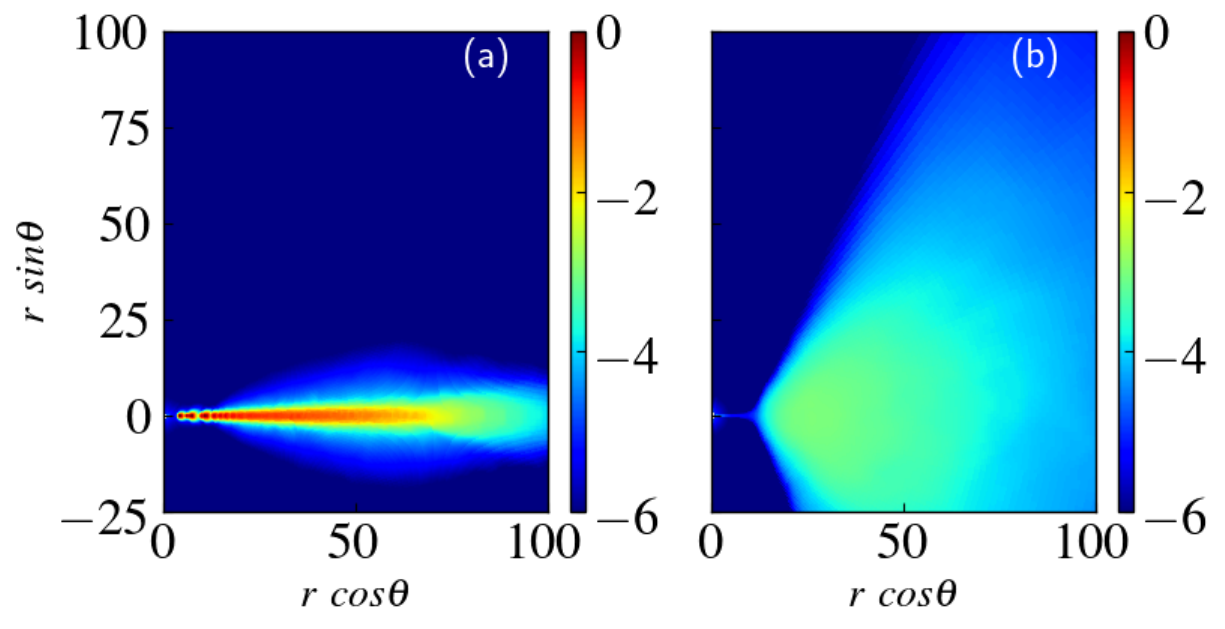}
    \caption{(a) Average density 
    snapshot without injected matter in between simulation time $t=4.5-5.0\times10^4t_g$. (b) Average density 
    snapshot without the initial thin disk in between simulation time $t=4.5-5.0\times10^4t_g$.}
    \label{fig:apx-1}
\end{figure}
To assess the robustness of our initial setup, we ran two simulation setups: one without any injected matter and another with only injected matter, excluding the initial thin disk. The average density profiles in between $t=4.5-5.0
\times 10^4$ is shown in panels (a) and (b) of Fig.~\ref{fig:apx-1}, respectively. This simulation time corresponds to $t\sim2.455$ $M_{\rm BH}/10\,M_\odot$ sec. Note that the result of the combination of the thin disk and the injected matter is presented in Fig.~\ref{fig:denisty}(b). 
Without magnetized injected matter, the thin disk can not transport angular momentum and maintain its thin disk structure throughout the simulation. On the other hand, without the thin disk, the injected matter forms a thick torus around the black hole, as shown in Fig.~\ref{fig:apx-1}(b). The size and shape of the torus are influenced by the value of $f_0$, which is $0.4$ in our test. This value determines the angular momentum of the injected matter. A higher $f_0$ value means the matter takes longer to reach the black hole, and the developed torus position becomes farther away.
If we kept injecting similar matter over a long period, radiative cooling might lead to the formation of a thin disk from the thick torus. This process represents the rising phase of an outburst in BH-XRBs. However, due to simulation time limits, we added an initial thin disk near the black hole to represent only the declining phase of the outburst in BH-XRBs. We plan to explore the full outburst cycle in future studies. We used a normalization factor of density $(0.1)$ in equation \ref{eq-rhoin}, ensuring that the injected density is much lower than the density of the thin disk.  Otherwise, the thin disk will be depleted very fast. With this constant, the density maximum for the torus in Fig.~\ref{fig:apx-1}(b) is three orders of magnitude lower than that of the density maximum of the thin disk in Fig.~\ref{fig:apx-1}(a). Our main goal is to show a qualitative representation of the outburst's decaying phase, so we kept this factor constant.

\section{Accretion rate with viscosity}
\begin{figure}
    \centering
    \includegraphics[scale=0.45]{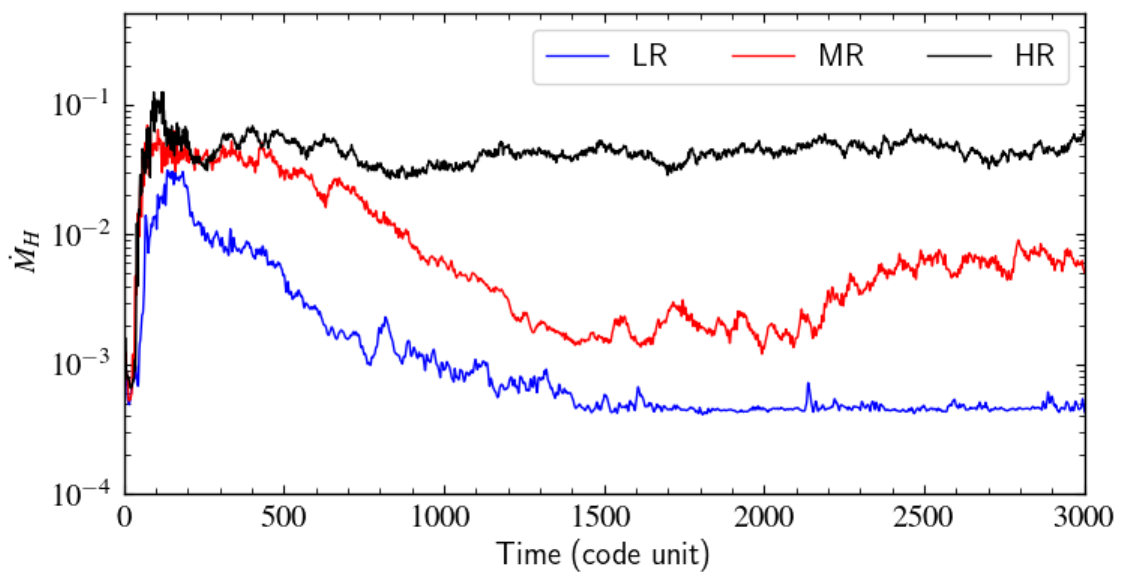}
    \caption{Evolution of accretion rate at the event horizon for model LR, MR, and HR.}
    \label{fig:apx-2}
\end{figure}

\begin{figure}
    \centering
    \includegraphics[scale=0.4]{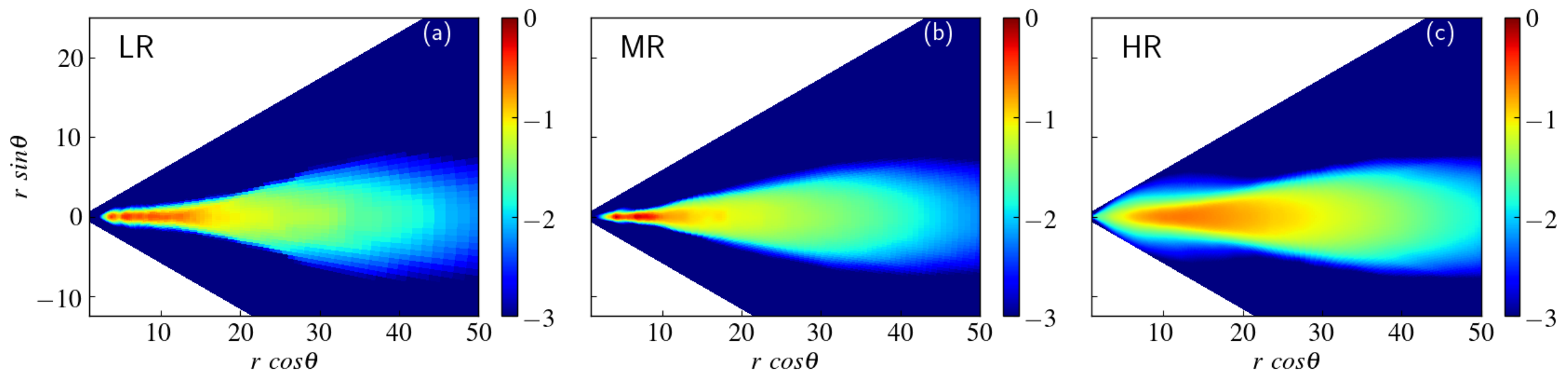}
    \caption{Time averaged density snapshot in the poloidal plane for model (a) LR, (b) MR, and (c) HR at simulation time $t=2000-3000t_g$.}
    \label{fig:apx-3}
\end{figure}

To study the effects of viscosity in the accretion disk, we consider three 3D runs of the thin-disk setup with different resolutions: 1. $160 \times 64 \times 64$ (LR) 2. $320 \times 128 \times 128$ with two refinement level(MR) and 3. $640 \times 256 \times 256$ with two refinement level (HR). 
For these runs, we put a weak loop magnetic field inside the thin disk with $\beta_{\rm min}=10$. Also, to reduce the extent of the thin disk by reducing the value of $r_{\rm out}=20$, which is $r_{\rm out}=30$ for the 2D models.
Since we only wanted to study the evolution of the disk and therefore, we simulate within $\theta=\pi/3$ to $\theta=2/3\pi$ and maximum resolution is concentrated within $\theta=\pi/2.4$ to $\theta=\pi - \pi/2.4$ (same as 2D models). Due to high numerical cost, we only perform simulation up to $t=3000t_g$. In Fig.~\ref{fig:apx-2} and \ref{fig:apx-3}, we compare the accretion rate at the event horizon $(\dot{M}_H)$ and the density profile for the three models (LR, MR, and HR). For these models, the average value of MRI quality factors ($<Q_\theta>,<Q_r>,<Q_\phi>$) are of the order of $\sim 1$(LR), $\sim 5-6$ (MR), and $\sim 10$ (HR) close to the black hole $r<10$, respectively. Note that to resolve MRI, the quality factor must be $Q\sim6$ or greater \citep[see][]{Sano-etal2004}. Accordingly, in the model LR, MR, and HR, the
MRI is under-resolved, marginally-resolved, and well-resolved, respectively. As a result, angular momentum transport due to MRI is the most efficient in model HR and inefficient in model LR. This result leads to two orders of magnitude higher accretion rates at the horizon for model HR as compared to LR, which can be clearly observed in Fig.~\ref{fig:apx-2}.

For completeness, we also show the time average density distribution for models LR(a), MR(b), and HR(c) within $t=2000-3000t_g$ in Fig.~\ref{fig:apx-3}.
By comparing these panels, we see that for model LR accretion disk is detached from the event horizon, which is quite similar to our 2D runs (Fig. \ref{fig:denisty}a). For higher resolution (HR), the accretion matter is attached up to the horizon, which supplies the black hole with a consistent accretion rate.

\label{lastpage}
\end{document}